\documentclass{nature}

\usepackage{cite}
\usepackage{amsmath,amssymb,amsfonts}
\usepackage[pdftex]{graphicx}
\usepackage{textcomp}
\usepackage{dcolumn}
\usepackage{bm}
\usepackage{comment}
\usepackage{multirow}
\usepackage{hhline}
\usepackage{ulem}
\usepackage{caption}
\usepackage{subcaption}

\title{Complementary junction field-effect transistor logic gate operational at 300$^\circ$C with 1.4 V supply voltage.}

\author{Mitsuaki Kaneko$^{1}$, Masashi Nakajima$^{1}$, Qimin Jin$^1$ \& Tsunenobu Kimoto$^1$}

\begin{document}

\maketitle

\begin{affiliations}
 \item The Department of Electronic Science and Engineering, Kyoto University, Kyoto 615-8510, Japan
\end{affiliations}

\begin{abstract}

Integrated circuits (ICs) that can operate at high temperature have a wide variety of applications in the fields of automotive, aerospace, space exploration, and deep-well drilling. 
Conventional silicon-based complementary metal-oxide-semiconductor (CMOS) circuits cannot work at higher than 200 $^\circ$C, leading to the use of wide bandgap semiconductor, especially silicon carbide (SiC).
However, high-density defects at an oxide-SiC interface make it impossible to predict electrical characteristics of SiC CMOS logic gates in a wide temperature range and high supply voltage (typically $\mathbf{\geqq 15}$ V) is required to compensate their large logic threshold voltage shift.
Here, we show that SiC complementary logic gates composed of p- and n-channel junction field-effect transistors (JFETs) operate at 300 $^\circ$C with a supply voltage as low as 1.4 V.
The logic threshold voltage shift of the complementary JFET (CJFET) inverter is 0.2 V from room temperature to 300 $^\circ$C.
Furthermore, temperature dependencies of the static and dynamic characteristics of the CJFET inverter are well explained by a simple analytical model of SiC JFETs.
This allows us to perform electronic circuit simulation, leading to superior designability of complex circuits or memories based on SiC CJFET technology, which operate within a wide temperature range.

\end{abstract}


\clearpage

The electronics operational at high temperature has a wide variety of applications in the fields of automotive, aerospace, and energy industry.\cite{French:0cjba,Watson:0hpba} 
However, conventional silicon (Si) complementary metal-oxide-semiconductor (CMOS) ICs are not operational at higher than 200 $^\circ$C since p-n junctions do not work as device isolation.
Then, Si CMOS on silicon-on-insulator (SOI) is developed and push the temperature limitation up to 300 $^\circ$C.\cite{Jeon:0uqba}
For further high temperature operation, the use of a wide bandgap semiconductor has attracted much attention.\cite{Neudeck:2002it}
The intrinsic carrier density in a wide bandgap semiconductor is many-orders-of-magnitude lower than that in Si, where p-n junction can be used as device isolation at higher temperature. 
Among wide bandgap semiconductors, silicon carbide (SiC) has several advantages in terms of fabrication of ICs such as mature ion implantation technology and mass production of SiC power devices.\cite{Kimoto:2016cc,Kimoto:2014ukba}
As in the case of Si, SiC CMOS circuits have been developed and high temperature operation has been demonstrated.\cite{Ryu:mmH2oSryba,Kuhns:0hxba}
However, SiC CMOS characteristics cannot be predicted within a wide temperature range due to the high-density defects at the silicon dioxide(SiO$_2$)/SiC interface,\cite{Potbhare:0fxba,(null):0jgba,Takeda:0ebba} leading to a high supply voltage ($V_{\rm dd}$, typically 15 V) and large power consumption.\cite{(null):0ikba}
Moreover, reliability of SiO$_2$ at high temperature is a serious concern since the high $V_{\rm dd}$ induces the high electric field inside SiO$_2$, resulting in the increase of gate leakage current.\cite{Le-Huu:0fkba}
Circuits composed of SiC n-channel junction field-effect transistors (n-JFETs) and resistors (JFET-R) were also proposed and long-term reliability has been proved.\cite{Neudeck:2009bzbaca,Neudeck:0coba}
However, their power dissipation especially in digital logic gates is large because of the use of depletion-mode JFETs and large $V_{\rm dd}$.

A complementary circuit can be assembled with p- and n-channel JFETs and a complementary JFET (CJFET) circuit should exhibit extremely low static power dissipation in the same manner as a CMOS circuit.
Operation of CJFET has been reported with gallium arsenide (GaAs)\cite{Zuleeg:VFqXNyk9ba, Zuleeg:bPyM2EVwba} and Si\cite{Kapoor:2010irba} for the development of ICs with ultra-low power loss. 
In CJFET circuits, however, $V_{\rm dd}$ has to be less than a built-in potential of a gate p-n junction to avoid current flowing through the gate p-n junction, indicating less flexibility of device design with Si and GaAs, where a typical value of a built-in potential is around 0.8-1.2 eV.
On the other hand, a gate p-n junction of SiC has a large built-in potential of about 3.0 eV at room temperature (RT), resulting in a reasonably higher $V_{\rm dd}$ and large noise margin.
Therefore, SiC CJFET technology has a potential to be a suitable choice as low-power-loss and high-reliability ICs operational at high temperature.

For fabrication of CJFET circuits, enhancement-mode JFETs are required and design of threshold voltage in both p- and n-JFETs ($V_{\rm Tp}$ and $V_{\rm Tn}$) is of importance.
Conventional SiC JFETs have used an epitaxial layer as a channel, leading to inevitable distribution of $V_{\rm Tn}$ inside a wafer due to the growth rate and doping distribution.\cite{Neudeck:2015boba}
To overcome the problem, we have proposed that all the device regions were fabricated by ion implantation and fabrication of enhancement-mode p- and n-JFETs on a same substrate was demonstrated.\cite{Kimoto:2018cz,Nakajima:2019kb}
In this article, SiC CJFET logic gates are assembled by the ion-implantation-based p- and n-JFETs.
Inverter operation at 300 $^\circ$C is demonstrated with $V_{\rm dd}$ of 1.4 V and its static and dynamic characteristics are well reproduced by a simple analytical model of SiC JFETs.
We have also demonstrated NOR operation at 300 $^\circ$C.

\section*{Fabrication of complementary JFET inverter}

\begin{figure}[p]
    \centering
    \includegraphics[bb = 0 0 478 744, width=120 mm]{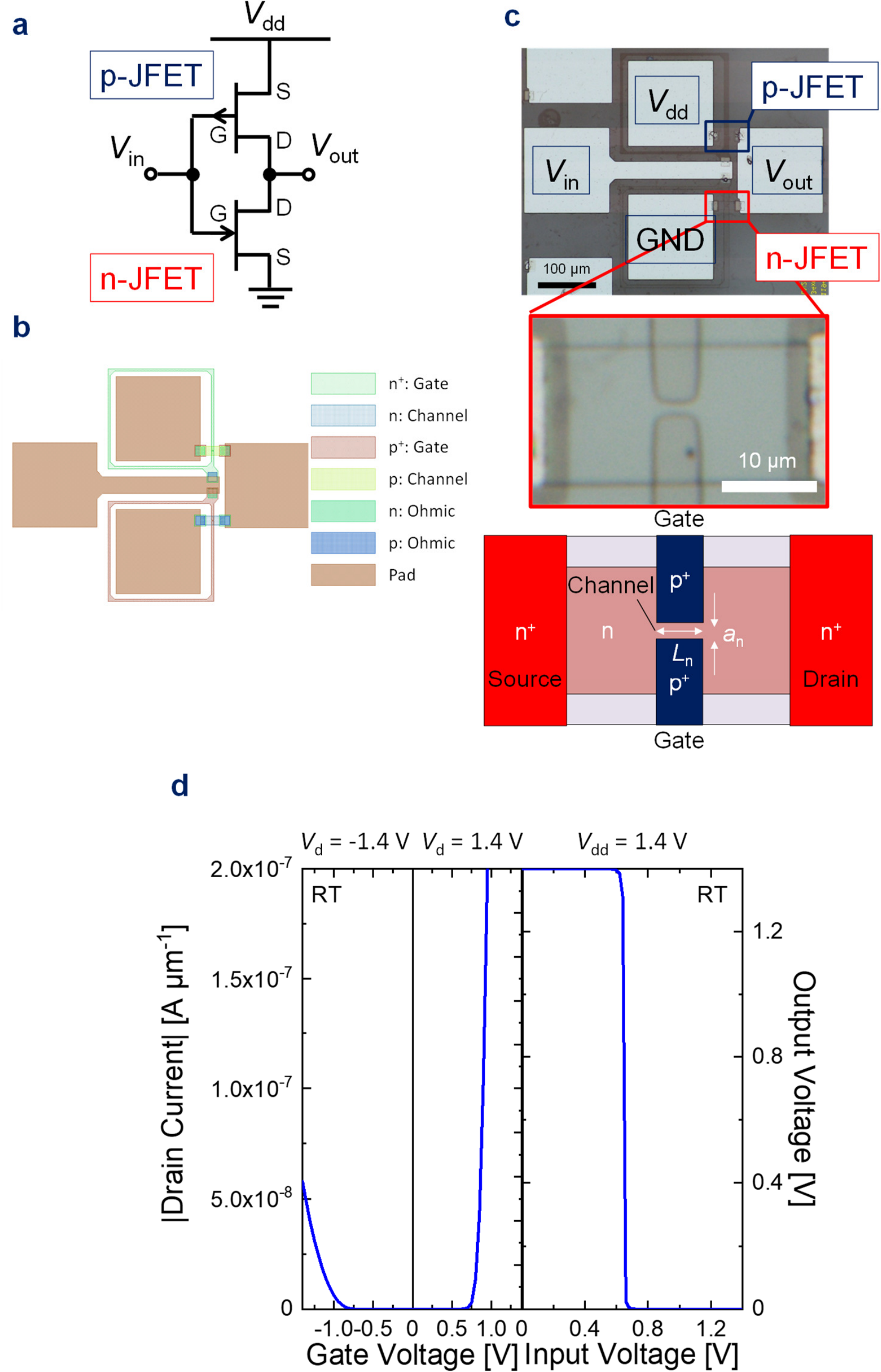}
    \caption{
    {\bf CJFET inverter.} {\bf a}, A CJFET inverter circuit diagram. {\bf b}, The layout design of the CJFET inverter. {\bf c}, An optical image of the fabricated CJFET inverter. The inset shows the magnified image of the n-JFET. {\bf d}, The transfer characteristics of the p- and n-JFETs and the VTC at room temperature.
    }
    \label{inv}
     \end{figure}

Figure \ref{inv}a shows a CJFET inverter circuit diagram.
A CJFET circuit can be assembled by replacing p- and n-MOS field-effect transistors (FETs) in a CMOS circuit by p- and n-JFETs, respectively.
The layout design of a CJFET inverter and optical image of the fabricated CJFET inverter are shown in Fig. \ref{inv}b,c.
The CJFET inverter was fabricated by ion implantation into a 4H-SiC high-purity semi-insulating substrate.
The semi-insulating region is expected to work as device isolation.
Total ion implantation steps are four and details of ion implantation conditions are provided in Supplementary Table \ref{tab11}.
As shown in the inset of Fig. \ref{inv}c, the channel of the JFET is controlled by side gates, which enhances transconductance compared to a standard single gate JFET.
The channel width ($W_{\rm p}$ or $W_{\rm n}$ in the p- or n-JFET) is defined as the direction orthogonal to the sample surface and $W_{\rm p}$ and $W_{\rm n}$ were estimated as 0.6 and 0.45 $\mu$m by secondary ion mass spectrometry.
The channel thickness ($a_{\rm p}$ or $a_{\rm n}$), which determines the $V_{\rm Tp}$ or $V_{\rm Tn}$, and the channel length ($L_{\rm p}$ or $L_{\rm n}$) were designed by the mask layout for ion implantation.
Owing to the lateral channeling of implanted atoms, the actual values of $a_{\rm p}$, $a_{\rm n}$, $L_{\rm p}$, and $L_{\rm n}$ differ from the mask design,\cite{JinJin} which will be estimated from the electrical characteristics below.

The transfer characteristics of the fabricated p- and n-JFETs assembling the CJFET inverter (Fig. \ref{inv}c) and the voltage transfer characteristic (VTC) of the CJFET inverter at RT are shown in Fig. \ref{inv}d.
Although the p- and n-JFETs have the similar channel thickness and width, the n-JFET shows higher drain current since electron mobility is larger than hole mobility in SiC and the ionization ratio of Al in the p-channel is low.
For the analysis of the CJFET circuits, a simple analytic model of a JFET was applied, the details of which are described in Supplementary Note 1.
From the transfer characteristics of the p- and n-JFETs, $V_{\rm Tp}$ and $V_{\rm Tn}$ are determined as 0.63 and 0.61 V, leading to the values of $a_{\rm p}$ and $a_{\rm n}$ as 460 and 458 nm, respectively.
Although the edges of the side gates are rounded due to the lateral straggling, 
$L_{\rm p}$, and $L_{\rm n}$ are not strongly modified and estimated as 4 $\mu$m, which are the same values of the mask design.
The logic threshold voltage ($V_{\rm th}$) of the CJFET is 0.65 V with $V_{\rm dd}$ of 1.4 V (Fig. \ref{inv}d).
As discussed above, SiC has a wide bandgap, leading to a large built-in potential between a p-n junction.
Then, forward current does not flow at the gate p-n junctions in the p- and n-JFETs even with $V_{\rm dd}$ of 1.4 V at RT.

\section*{Static characteristics}

\begin{figure}[p]
        \centering
        \includegraphics[bb = 0 0 496 505, width=150 mm]{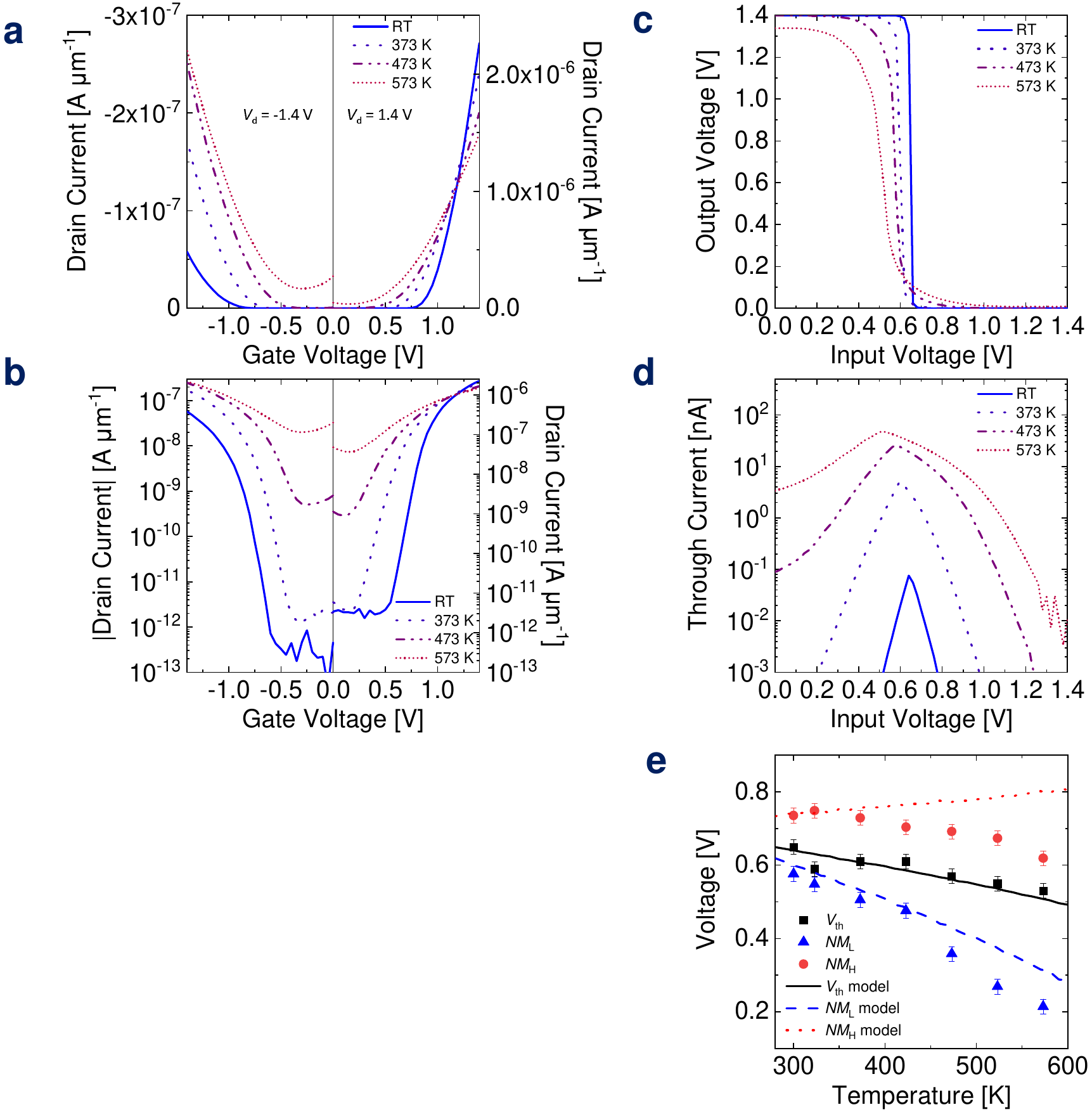}
        \caption{
        {\bf Static characteristics of the CJFET inverter.} {\bf a,b}, The transfer characteristics of the p- and n-JFETs from RT to 573 K in linear (a) and logarithmic (b) scales. {\bf c,d}, The temperature dependence of the VTC (c) and through current (d). The gate leakage current is subtracted from the through current. {\bf e}, The temperature dependence of the $V_{\rm th}$, $NM_{\rm L}$, and $NM_{\rm H}$. The symbols denote the experimental data. Error bars mean the voltage step in the measurements. The solid, dashed, and dotted lines are the $V_{\rm th}$, $NM_{\rm L}$, and $NM_{\rm H}$ extracted from the analytical model.
        }
        \label{temp}
    \end{figure}

The temperature dependence of the transfer characteristics in the p- and n-JFETs is depicted in Fig. \ref{temp}a,b.
The measurement temperature range is from RT to 573 K.
The drain current of the p-JFET is enhanced with elevating the temperature since the hole density in the p-channel increases by enhancement of dopant ionization.
The decrease in the drain current of the n-JFET is attributed to electron mobility lowering.
These trends are the same as the side gate p- and n-JFETs separately fabricated in our previous study.\cite{Nakajima:2019kb}
On the other hand, the leakage current at a gate voltage of 0 V in the fabricated p- and n-JFETs dramatically increases with elevating the temperature. 
The increasing rate against temperature is much larger than those in our previously reported JFETs (see Supplementary Fig. 1).
The leakage current in the JFETs assembling the CJFET inverter mainly consists of the gate-source leakage current whereas the gate leakage current was negligible even at 573 K in the separately fabricated JFETs.
In the CJFET inverter, parasitic p-i-p (the source region in the p-JFET to the gate region in the n-JFET) or n-i-n (the source region in the n-JFET to the gate region in the p-JFET) structures exist (see Fig. \ref{inv}b,c). 
Since the resistivity of semi-insulating region decreases with increasing the temperature,\cite{Muzykov:0tgba} such parasitic p-i-p or n-i-n structure may contribute to the gate leakage current.
In this study, we have not fabricated a metal interconnect layer and the gate region had to be large enough to make the side gate structure, leading to an increase in the area of the parasitic p-i-p or n-i-n structure.
Such leakage current is likely suppressed by shrinking the area of the parasitic p-i-p or n-i-n structure or fabricating p- or n-well structure like conventional CMOS circuits.

The VTCs in the CJFET inverter within the temperature range between RT and 573 K are presented in Fig. \ref{temp}c.
The $V_{\rm th}$ shift from RT to 573 K is about 0.2V.
The output voltage with the low input voltage (0 V) at 573 K is slightly lower than the $V_{\rm dd}$ (1.4 V), which is ascribed to the gate leakage current discussed above.
The through current subtracting the gate leakage current from RT to 573 K is shown in Fig. \ref{temp}d.
With elevating the temperature, the absolute values of $V_{\rm Tp}$ and $V_{\rm Tn}$ become smaller, leading to expansion of the voltage range where the through current flows and the higher peak current at higher temperature.
Nevertheless, the highest through current is limited as small as about 50 nA.
The static current at the input voltage of low (0 V) and high (1.4 V) is negligibly small ($< 0.1$ nA) at up to the temperature of 473 K, indicating the extremely low static power dissipation.
At 573 K, small through current is observed with the input voltage of 0 V even after subtracting the gate leakage current.
From the source region of the p-JFET to the source region of the n-JFET, parasitic p-i-n diode exists and the diode was forward biased by applying $V_{\rm dd}$.
Then, small current flew through the parasitic p-i-n diode since the built-in potential of the p-i-n diode was also lowered by elevating the temperature, which can also be prevented by changing the device layout as discussed in the gate leakage current.

The temperature dependencies of low and high noise margins ($NM_{\rm L}$ and $NM_{\rm H}$) of the CJFET inverter are shown in Fig. \ref{temp}e.
Definition of $NM_{\rm L}$ and $NM_{\rm H}$ is given in Supplementary Note 1.
Both of the $NM_{\rm L}$ and $NM_{\rm H}$ in the fabricated CJFET become smaller with elevating the temperature since the voltage range in the transition region becomes wider (see Fig. \ref{temp}c).
$NM_{\rm L}$ at 573 K (0.2 V) still remains much higher than the thermal voltage (0.049 eV), which avoids unintentional transition due to the thermal noise.
The dashed and dotted lines denote the $NM_{\rm L}$ and $NM_{\rm H}$ extracted from the analytical model (Supplementary Note 1).
These curves agree well with the experimental results up to about 423 K and gradually deviate with elevating the temperature.
Here, the subthreshold current of JFETs were not considered in the analytical model and non-ideal leakage current also flew as discussed above, resulting in the wider transition region and smaller $V_{\rm OH}$ in the fabricated CJFET inverter (see Supplementary Fig. 2).
Nevertheless, $V_{\rm th}$ extracted from the analytical model shows an excellent agreement with the experimental results within the measurement temperature range since the transition region does not affect much on the values of $V_{\rm th}$.

\section*{Dynamic characteristics}

\begin{figure}[p]
        \centering
        \includegraphics[bb = 0 0 501 425, width=160 mm]{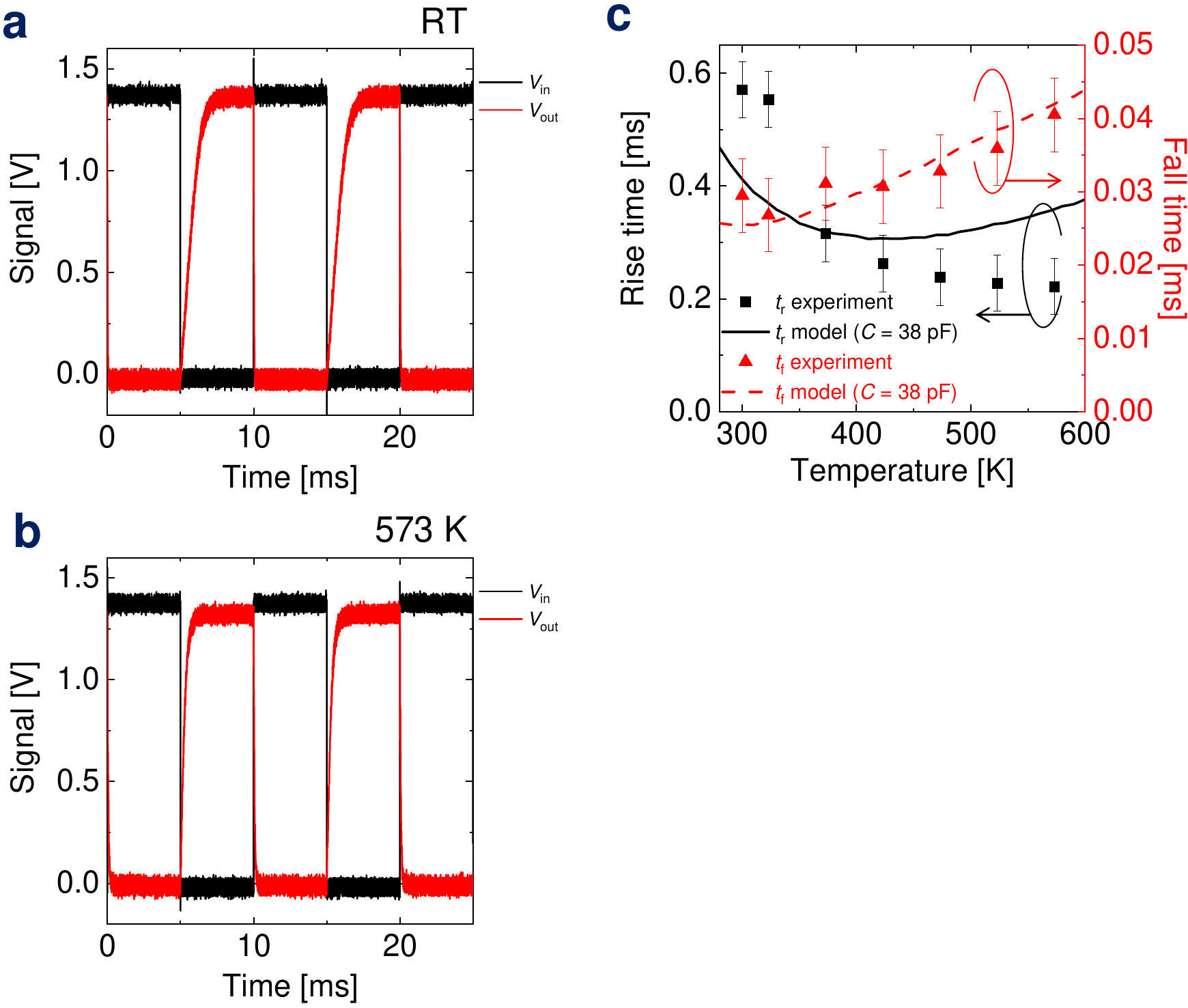}
        \caption{
        {\bf Dynamic characteristics of the CJFET inverter.} {\bf a,b}, Input voltage and output of the CJFET inverter at RT (a) and 573 K (b). {\bf c}, The temperature dependence of $t_{\rm r}$ and $t_{\rm f}$. The symbols denote the experimental data. Error bars represent noise height in oscilloscope. The solid and dashed lines are $t_{\rm r}$ and $t_{\rm f}$ obtained from the analytical model assuming a load capacitance of 38 pF.
        }
        \label{dyna}
    \end{figure}

Fig. \ref{dyna}a,b present dynamic characteristics of the CJFET inverter at RT and 573 K, respectively.
As demonstrated in the static characteristics (Fig. \ref{temp}), the output voltage level with $V_{\rm in}$ of 0 V at RT is equal to the $V_{\rm dd}$ (1.4 V) and slight lowering is observed at 573 K.
When focusing on the rise and fall times ($t_{\rm r}$ and $t_{\rm f}$) of the output level (defined as a time interval from 0\% or 100\% to 50\% level), the $t_{\rm f}$ is much shorter than the $t_{\rm r}$ since the drain current of the n-JFET (directly related to its $t_{\rm f}$) is larger than that of the p-JFET (its $t_{\rm r}$).
The $t_{\rm r}$ and $t_{\rm f}$ clearly depend on the temperature, which are plotted in Fig. \ref{dyna}c.
The monotonous decrease in the $t_{\rm r}$ with elevating the temperature is caused by the increased drain current in the p-JFET, due to the higher ionization ratio of aluminum acceptors and driving voltage enhancement by the $V_{\rm Tp}$ shift.
On the other hand, the $t_{\rm f}$ remains almost constant up to 423 K and gradually increases with temperature, which is attributed to competition between electron mobility lowering and higher driving voltage due to the $V_{\rm Tn}$ shift.

The $t_{\rm r}$ and $t_{\rm f}$ were calculated based on the analytical model (Supplementary Note 1).
Assuming that each JFET is biased under the saturation region within the time from 0\% or 100\% to 50\% output level, the $t_{\rm r}$ and $t_{\rm f}$ are expressed as the following formulae,
\begin{eqnarray}
    t_{\rm r} &=& \frac{CV_{\rm out\mathchar`-high}}{2I_{\rm dp}},\\
    t_{\rm f} &=& \frac{CV_{\rm out\mathchar`-high}}{2I_{\rm dn}},
\end{eqnarray}
where $C$ is the load capacitance and $I_{\rm dp}$ and $I_{\rm dn}$ are the drain current of the p- and n-JFETs, respectively.
The $V_{\rm out\mathchar`-high}$ corresponds to the $V_{\rm out}$ with $V_{\rm in}$ of 0 V (100\% output level).
The $t_{\rm r}$ and $t_{\rm f}$ were extracted from the Eqs. (1) and (2), which are depicted as the solid and dashed lines in Fig. 3c, respectively.
The $V_{\rm out\mathchar`-high}$ in Eqs. (1) and (2) is assumed to be $V_{\rm dd}$ in the whole temperature range.
We assembled a voltage follower circuit to measure the dynamic characteristics (see Measurement section) and $C$ is regarded as a fitting parameter, resulting in $C = 38$ pF.
The curves show a good agreement with the experiments.
The slight deviation between the calculation and experiments in $t_{\rm r}$ may be attributed to inaccuracy of the reported hole mobility in the p-JFET and unintentional voltage drop due to the leakage current discussed above.
Note that the long $t_{\rm r}$ and $t_{\rm f}$ originate from the small channel width, which can be improved by changing the device design such as a multi-channel structure.

\begin{figure}[p]
  \centering
  \includegraphics[bb = 0 0 531 362, width=\columnwidth]{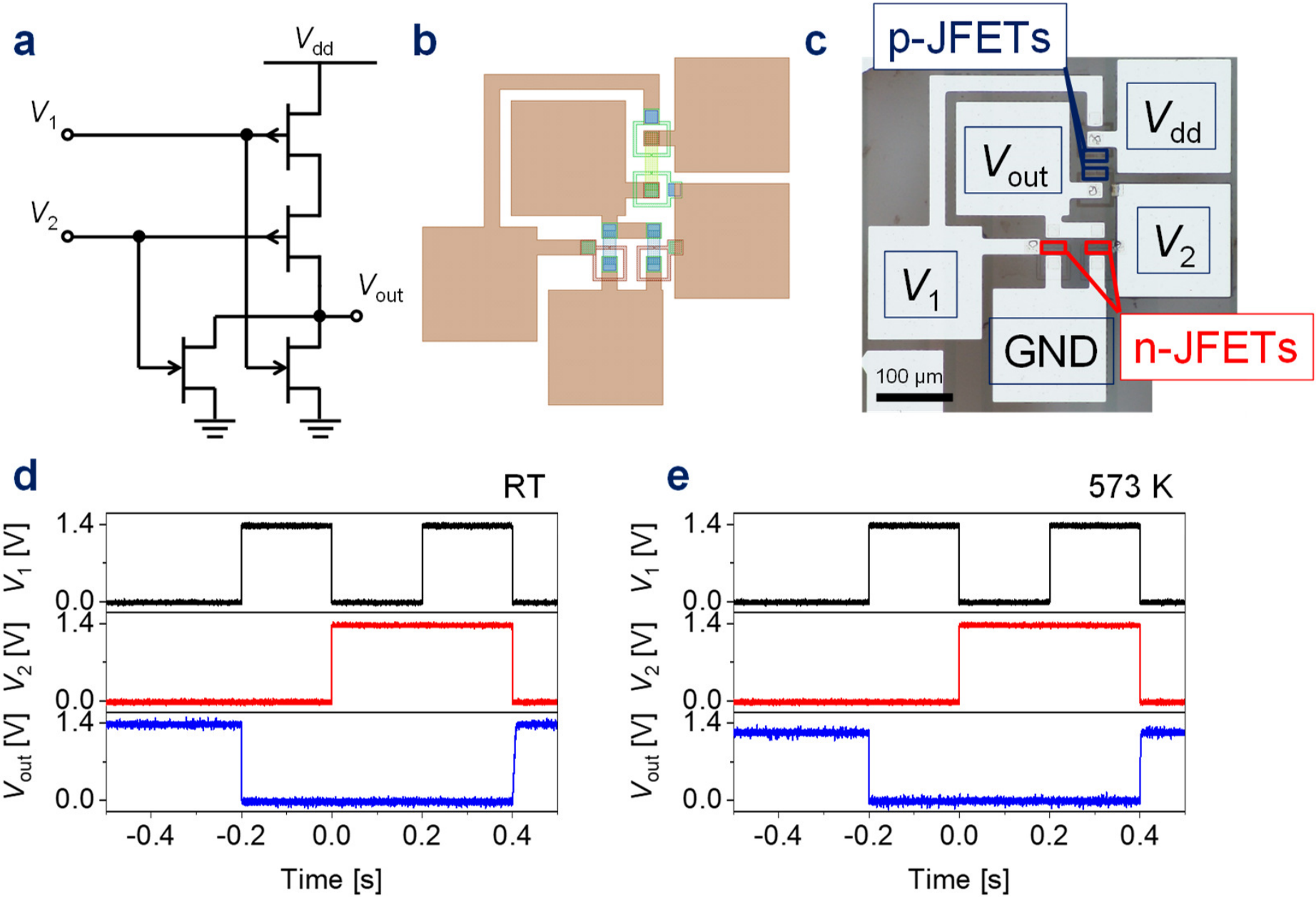}
    \caption{
        {\bf CJFET NOR operation.} {\bf a}, A CJFET NOR circuit diagram. {\bf b}, The layout design of the CJFET NOR logic gate. {\bf c}, An optical image of the fabricated CJFET NOR logic gate. {\bf d,e }, Input voltages and output of the CJFET NOR gate at RT (d) and 573 K (e).
    }
    \label{nor}
\end{figure}

Figures \ref{nor}a-c depict a CJFET NOR circuit.
As in the same as a CJFET inverter, a CJFET NOR circuit can be assembled in a replacement of MOSFETs in a CMOS circuit by JFETs.
Two input operations of the CJFET NOR circuit at RT and 573 K with $V_{\rm dd}$ of 1.4 V are demonstrated in Fig. \ref{nor}d,e.
Rail-to-rail (0--1.4 V) operation at RT and slight drop in $V_{\rm out}$ with $V_{\rm 1}$ and $V_{\rm 2}$ of 0 V at 573 K are confirmed, which are the same as can be seen in the CJFET inverter operation.

Here, we compare and discuss the characteristics of the reported high-temperature logic gates, which are listed in Table \ref{tab11}.
The SiC bipolar junction transistor (BJT) logic gates can operate with a single power supply and operation at higher than 500 $^\circ$C has been reported.\cite{Shakir:2019coba}
The SiC JFET-R logic gates also show high-temperature operation at higher than 800 $^{\circ}$C.\cite{Neudeck:2017gq}
Moreover, year-long stable operation of the JFET-R logic gates at 500 $^\circ$C has been achieved, which is of importance for the systems that cannot easily be repaired.\cite{Neudeck:2009un}
However, BJTs are current-controlled devices and the JFET-R logic gates use depletion-mode n-JFETs, leading to high static power consumption. 
On the other hand, the SiC CMOS logic gates show much smaller power consumption owing to complementary operation. 
However, high-density defects at the SiO$_2$/SiC interface make it difficult to control threshold voltages of p- and n-MOSFETs within a wide temperature range, requiring high $V_{\rm dd}$ (typically $\geqq 15$ V). 
Moreover, the temperature dependence of drain current in SiC p- and n-MOSFETs strongly depend on each fabrication process and it is rather difficult to develop a universally applicable analytical model for SiC MOSFETs, which is necessary for circuit simulations. 
Although the Si CMOS on SOI is well developed and already commercialized, its operational temperature is limited by the material properties of Si.
The SiC CJFET logic gates fabricated in this study work with a small and single power supply (1.4 V), leading to very small power consumption.
Since the SiC JFET-R logic gates demonstrate very high reliability, the SiC CJFET logic gates are also expected to be highly reliable, which should be tested in the future study.
Higher-temperature operation of SiC CJFET circuits is expected by improving the CJFET logic gate design.
The separately fabricated JFETs in our previous study show stable operation at 400 $^{\circ}$C,\cite{Nakajima:2019kb} meaning that the SiC CJFET can also operate at higher than 400 $^{\circ}$C. 

Gallium nitride (GaN) is another attractive wide bandgap semiconductor and a heterostructure of aluminum gallium nitride (AlGaN) and GaN makes two-dimensional electron gas (2DEG) at the interface, which is used as an n-channel in high electron mobility transistors (HEMTs).
Since p-type GaN formation by ion implantation is technologically difficult, logic circuits are basically composed of n-channel HEMTs.
$V_{\rm th}$ of AlGaN/GaN HEMTs can be modified by etching of AlGaN or re-growth of p-GaN under the gate region, which allows us to make depletion- and enhancement-mode HEMTs on the same substrate and assemble direct-coupled FET logic (DCFL) circuits.
Although 200-300 $^\circ$C operation of GaN DCFL circuits are reported,\cite{10.1016/j.sse.2008.09.001,10.1109/led.2017.2725908,10.1109/ted.2021.3075425} several issues should be addressed for higher temperature operation, such as leakage current from 2DEG to gate electrode, controllability and uniformity of $V_{\rm th}$, and device modeling including temperature dependencies.
Recently, several groups have reported fabrication of GaN p-channel FETs without ion implantation and assemble complementary integrated circuits,\cite{10.1109/led.2016.2515103,10.1109/led.2020.2987003,10.1109/led.2020.3039264} which dramatically reduces power consumption compared to GaN DCFL circuits, although the same issues of GaN DCFL circuits have to be solved for high-temperature operation.

\section*{Conclusion}

SiC CJFET logic gates were fabricated by ion implantation and inverter and NOR operations were demonstrated in a wide temperature range from RT to 573 K with a single and low supply voltage of 1.4 V.
The static characteristics, or $V_{\rm th}$, $NM_{\rm L}$, and $NM_{\rm H}$, and dynamic characteristics, or $t_{\rm r}$ and $t_{\rm f}$, were well explained by a simple analytical model, which indicates that electronic circuit simulation on SiC CJFETs can predict actual circuit operation.
In a similar manner of prosperity in the Si CMOS technology, SiC CJFET technology is widely applicable to digital and analog circuits and memories operating at high temperature with small power consumption.

\begin{methods}

\subsection{Sample preparation.}

A $4{}^\circ$-off-axis high-purity semi-insulating 4H-SiC(0001) substrate grown by high-temperature chemical vapor deposition was used as a starting material.
All the n-type and p-type regions were formed by ion implantation of phosphorus and aluminum, respectively.
A device-isolation process such as mesa etching was not performed in this study.
Silicon dioxide (SiO$_2$) with a thickness of 2-3 ${\rm \mu}$m was deposited as an ion implantation mask.
Patterning of the SiO$_2$ mask was performed using photolithography and dry etching.
The high-dose regions (n$^+$ and p$^+$) and low-dose regions (n and p) were implanted at 300 $^\circ$C and room temperature (RT), respectively.
The implantation conditions are summarized in Supplementary Table 1.
The doping concentration of the gate and channel regions were $5 \times 10^{19}$ and $5 \times 10^{16}$ cm$^{-3}$, respectively, which were confirmed by secondary ion mass spectrometry.
After the ion implantation, activation annealing was conducted at 1650 $^\circ$C for 10 min.
Ti/Al and Ni were thermally evaporated as ohmic contacts for p- and n-type regions, respectively.
1-${\rm \mu}$m-thick Al was finally deposited as metal pads.
A detailed device layout is shown in Fig. 1b.

\subsection{Measurements.}
All the measurements were performed under vacuum condition using a probe station with a temperature controller stage (MJ-10-P6K, APPOLO WAVE).
The sample temperature was monitored with a thermocouple sensor located close to the sample.
The static characteristics were obtained with a Keithley 4200A semiconductor parameter analyzer.
The input signal for the dynamic characteristics was generated by a signal generator (81160A, Agilent Technologies).
Due to the high output impedance of the CJFET logic gates in this study, a voltage follower circuit assembled with an op-amp (LT1793, Linear Technology) was connected to the output of the CJFET logic gates.
Then, an oscilloscope (DSO9254A, Agilent Technologies) was used to obtain the output signals.

\subsection{Data availability.}
The data within this paper are available from the corresponding author upon reasonable request.

\end{methods}

\clearpage

\section*{References}

\bibliography{ful,add}

\begin{thebibliography}{1}
\bibitem{dummy} Articles are restricted to 50 references, Letters
to 30.
\end{thebibliography}
\bibliographystyle{naturemag}

\begin{addendum}
 \item[Acknowledgements] This work was supported by the Program an Open Innovation Platform with Enterprises, Research Institute and Academia (OPERA) from the Japanese Science and Technology Agency, a research granted from the Murata Science Foundation, and the Inamori Foundation.
 \item[Author contributions] T.K. supervised the entire project with M.K. M.N. designed the devices with M.K. M.N. fabricated the sample with Q.J. and M.K. M.K. performed the experiments and analyzed the data with M.N. M.K. wrote the manuscript with T.K. All authors discussed the results and contributed to the manuscript.
 \item[Competing Interests] The authors declare that they have no
competing financial interests.
 \item[Correspondence] Correspondence and requests for materials
should be addressed to M.K. or T.K. \\(email: kaneko@semicon.kuee.kyoto-u.ac.jp; kimoto@kuee.kyoto-u.ac.jp).
\end{addendum}

\clearpage

\begin{table*}
  \begin{center}
  \caption{Comparison of the high-temperature logic gates.}
  \label{tab11}
  \begin{tabular}{ c c c c c c c }
  \hline
 &\multirow{2}{*}{$V_{\rm dd}$}  & \multirow{2}{*}{$V_{\rm ss}$}& $V_{\rm th}$ & Power & \multirow{2}{*}{Reliability}   & $T$ \\
 & &  & shift & consumption & & (demonstrated) \\
  \hline
SiC BJT               &$\geqq 15$ V&GND     &small     & high & high          & $\geqq 500$ $^{\circ}$C   \\
SiC JFET-R            &$\geqq 25$ V&$\leqq-25$ V&small & high & very high     & $\geqq 800$ $^{\circ}$C \\
SiC CMOS              &$\geqq 15$ V&GND     &medium    & low       & SiO$_{\rm 2}$-limited & $\geqq 400$ $^{\circ}$C       \\
Si CMOS on SOI        &1-5 V       &GND     &small     & very low  & SiO$_{\rm 2}$-limited & $\leqq 300$ $^{\circ}$C \\
GaN DCFL              &$\geqq 3$ V &GND     &medium    & high      & $-$ & $\leqq 350$ $^{\circ}$C \\
\hline
SiC CJFET (this study)&$\leqq 2$  V&GND     &small & very low  & high (expected) & $ 300$ $^{\circ}$C \\
\hline
\end{tabular}
  \end{center}
  \end{table*}


\clearpage

\section*{Supplementary Note 1}
An analytical model used in this paper to describe electrical characteristics of SiC p- and n-JFETs is introduced.
Drain current of p- and n-JFETs ($I_{\rm{dp}}$ and $I_{\rm{dn}}$) with threshold voltages of $V_{\rm Tp}$ and $V_{\rm Tn}$ for uniformly doped p- and n-channel regions is presented as the following formulae under gradual-channel approximation,\cite{Sze1}

\begin{align}
        I_{\rm{dp}} &= 
          \begin{cases}
            0 & (\text{$V_{\rm gp} < V_{\rm Tn}$})\\
            -G_{\rm{ip}} \left[ |V_{\rm{dp}}|-\frac{2}{3}\psi_{\rm{pp}} \left\{{\left(\frac{\psi_{\rm{jp}}-|V_{\rm{gp}}-V_{\rm{dp}}|}{\psi_{\rm{pp}}}\right)}^{1.5}-{\left(\frac{\psi_{\rm{jp}}-|V_{\rm{gp}}|}{\psi_{\rm{pp}}}\right)}^{1.5} \right\}  \right] & (\text{$V_{\rm gp} > V_{\rm Tp}$, \text{$V_{\rm dp} < V_{\rm gp} - V_{\rm Tp}$}})\\
            -G_{\rm{ip}} \left[ \frac{\psi_{\rm{pp}}}{3}-(\psi_{\rm{jp}}-|V_{\rm{gp}}|) \left( 1-\frac{2}{3}\sqrt{\frac{\psi_{\rm{jp}}-|V_{\rm{gp}}|}{\psi_{\rm{pp}}}} \right) \right] & (\text{$V_{\rm gp} > V_{\rm Tp}$, \text{$V_{\rm dp} > V_{\rm gp} - V_{\rm Tp}$}})\\
          \end{cases}
          \tag{S1}
          \label{eq:pcaldrain}
        \end{align}
\begin{align}
        I_{\rm{dn}} &= 
          \begin{cases}
            0 & (\text{$V_{\rm gn} < V_{\rm Tn}$})\\
            G_{\rm{in}} \left[ V_{\rm{dn}}-\frac{2}{3}\psi_{\rm{pn}} \left\{{\left(\frac{\psi_{\rm{jn}}-V_{\rm{gn}}+V_{\rm{dn}}}{\psi_{\rm{pn}}}\right)}^{1.5}-{\left(\frac{\psi_{\rm{jn}}-V_{\rm{gn}}}{\psi_{\rm{pn}}}\right)}^{1.5} \right\}  \right] & (\text{$V_{\rm gn} > V_{\rm Tn}$, \text{$V_{\rm dn} < V_{\rm gn} - V_{\rm Tn}$}})\\
            G_{\rm{in}} \left[ \frac{\psi_{\rm{pn}}}{3}-(\psi_{\rm{jn}}-V_{\rm{gn}}) \left( 1-\frac{2}{3}\sqrt{\frac{\psi_{\rm{jn}}-V_{\rm{gn}}}{\psi_{\rm{pn}}}} \right) \right] & (\text{$V_{\rm gn} > V_{\rm Tn}$, \text{$V_{\rm dn} > V_{\rm gn} - V_{\rm Tn}$}}) \\
          \end{cases}
          \tag{S2}
          \label{eq:ncaldrain}
        \end{align}
where $\psi_{\rm jp}$ ($\psi_{\rm jn}$) is the built-in potential of the p-n junction between the gate and channel region in a p-JFET (n-JFET). 
$\psi_{\rm{pp}}$ ($\psi_{\rm{pn}}$) is the pinch-off potential of a p-JFET (n-JFET).
$G_{\rm{ip}}$ and $G_{\rm{in}}$ are expressed with the following formulae, $G_{\rm{ip}}$ = $\frac{q{W_{\rm{p}}}{\mu_{\rm{p}}}p_{\rm p}a_{\rm{p}}}{L_{\rm{p}}}$ and $G_{\rm{in}}$ = $\frac{q{W_{\rm{n}}}{\mu_{\rm{n}}}n_{\rm n}a_{\rm{n}}}{L_{\rm{n}}}$.
$q$ is the elementary charge.
$W_{\rm{p}}$ ($W_{\rm{n}}$), $L_{\rm{p}}$ ($L_{\rm{n}}$), and $a_{\rm p}$ ($a_{\rm n}$) are the channel width, length and thickness of a p-JFET (n-JFET), respectively.
$\mu_{\rm{p}}$ ($\mu_{\rm{n}}$) and $p_{\rm p}$ ($n_{\rm n}$) are the hole (electron) mobility and density in the channel region.

Voltage transfer characteristics (VTCs) of a SiC CJFET inverter were calculated using Eqs. (\ref{eq:pcaldrain}) and (\ref{eq:ncaldrain}).
As indicated in the inverter circuit diagram (Fig. 1), the supply voltage ($V_{\rm dd}$) and input voltage ($V_{\rm in}$) are substituted for $V_{\rm gp}$ and $V_{\rm gn}$ as following, $V_{\rm gp} = V_{\rm in} - V_{\rm dd}$ and $V_{\rm gn} = V_{\rm in}$.
Then, $V_{\rm out}$ is determined as the $V_{\rm{dp}}$ ($= V_{\rm{dn}}$) where $I_{\rm{dp}} = I_{\rm{dn}}$.
For the calculation of Eqs. (\ref{eq:pcaldrain}) and (\ref{eq:ncaldrain}), material properties of 4H-SiC are used, which are obtained from epitaxially grown p- and n-type epilayers.

Next, $V_{\rm th}$, $NM_{\rm L}$, and $NM_{\rm H}$ are calculated.
Eqs. (\ref{eq:pcaldrain}) and (\ref{eq:ncaldrain}) are approximated with the following functions when the gate voltage is close to the threshold voltage of JFETs,

\begin{align}
    I_{\rm{dp}} &= 
      \begin{cases}
        0 & (\text{$V_{\rm gp} < V_{\rm Tn}$})\\
       \frac{G_{\rm{ip}}}{2\psi_{\rm{pp}}}(V_{\rm gp}-V_{\rm Tp})V_{\rm dp} & (\text{$V_{\rm gn} > V_{\rm Tn}$, \text{$V_{\rm dn} < V_{\rm gn} - V_{\rm Tn}$}})\\
         \frac{G_{\rm{ip}}}{4\psi_{\rm{pp}}}(V_{\rm gp}-V_{\rm Tp})^2 & (\text{$V_{\rm gp} > V_{\rm Tp}$, \text{$V_{\rm dp} > V_{\rm gp} - V_{\rm Tp}$}})\\
      \end{cases}
      \tag{S3}
      \label{eq:pdrain_app2}
    \end{align}
    
    \begin{align}
    I_{\rm{dn}} &= 
      \begin{cases}
        0 & (\text{$V_{\rm gn} < V_{\rm Tn}$})\\
        \frac{G_{\rm{in}}}{2\psi_{\rm{pn}}}(V_{\rm gn}-V_{\rm Tn})V_{\rm dn} & (\text{$V_{\rm gn} > V_{\rm Tn}$, \text{$V_{\rm dn} < V_{\rm gn} - V_{\rm Tn}$}})\\
        \frac{G_{\rm{in}}}{4\psi_{\rm{pn}}}(V_{\rm gn}-V_{\rm Tn})^2 & (\text{$V_{\rm gn} > V_{\rm Tn}$, \text{$V_{\rm dn} > V_{\rm gn} - V_{\rm Tn}$}})\\
      \end{cases}
      \tag{S4}
      \label{eq:ndrain_app2}
    \end{align}

When a CJFET inverter reaches $V_{\rm th}$, both of the p- and n-JFETs operate under saturation regions.
Therefore, $V_{\rm th}$ corresponds to $V_{\rm in}$ which satisfies the following equation,

\begin{equation}
    \frac{G_{\rm{ip}}}{4\psi_{\rm{pp}}}(V_{\rm in}-V_{\rm dd}-V_{\rm Tp})^2 = \frac{G_{\rm{in}}}{4\psi_{\rm{pn}}}(V_{\rm in}-V_{\rm Tn})^2,
    \tag{S5}
   \label{eq:Vinv}
   \end{equation}
leading to
    \begin{equation}
        V_{\rm th} = \frac{V_{\rm dd}+V_{\rm Tp}+\sqrt{\beta{\rm _R}}V_{\rm Tn}}{1+\sqrt{\beta{\rm _R}}},
        \tag{S6}
       \label{eq:Vinv2}
    \end{equation}
where $\beta{\rm _R}$ is defined as $\beta{\rm _R} = \frac{G_{\rm{in}}\psi_{\rm{pp}}}{G_{\rm{ip}}\psi_{\rm{pn}}} $.

When defining $V_{\rm IL}$ and $V_{\rm OH}$ as $V_{\rm in}$ and $V_{\rm out}$ at ${\rm d}V_{\rm out}/{\rm d}V_{\rm in} = -1$ within a transition region in a VTC from the high level to $V_{\rm th}$ and $V_{\rm IH}$ and $V_{\rm OL}$ as $V_{\rm in}$ and $V_{\rm out}$ at ${\rm d}V_{\rm out}/{\rm d}V_{\rm in} = -1$ within a transition region in a VTC from $V_{\rm th}$ to the low level, $NM_{\rm L}$ and $NM_{\rm H}$ are expressed as the following formulae,
\begin{equation}
    NM_{\rm L} = V_{\rm IL} - V_{\rm OL},  
    \tag{S7}
\end{equation}
\begin{equation}
    NM_{\rm H} = V_{\rm OH} - V_{\rm IH}.  
    \tag{S8}
\end{equation}
When $V_{\rm in} = V_{\rm IL}$, the p- and n-JFETs work under non-saturation and saturation regions, respectively, where the following function holds,
\begin{equation}
    \frac{G_{\rm{ip}}}{2\psi_{\rm{pp}}}(V_{\rm in}-V_{\rm dd}-V_{\rm Tp})(V_{\rm out}-V_{\rm dd}) = \frac{G_{\rm{in}}}{4\psi_{\rm{pn}}}(V_{\rm in}-V_{\rm Tn})^2.
    \tag{S9}
    \label{eq:NML1}
\end{equation}
By differentiating the above equation and substituting $V_{\rm in}$ = $V_{\rm IL}$ and $\frac{{\rm d}V{\rm _{out}}}{{\rm d}V{\rm _{in}}}$ = $-$1,
\begin{equation}
    V_{\rm out}-V_{\rm IL}+V_{\rm Tp} = \beta{\rm _R}(V_{\rm IL}-V_{\rm Tn})
    \tag{S10}
    \label{eq:NML2}
    \end{equation}
is obtained. 
$V_{\rm IL}$ is extracted by solving the Eqs. (\ref{eq:NML1}) and (\ref{eq:NML2}).
$V_{\rm OH}$ is obtained from the calculated VTC at $V_{\rm in}$ = $V_{\rm IL}$.
$V_{\rm IH}$ and $V_{\rm OL}$ can be extracted in a similar manner shown above.

\setcounter{figure}{0}
\begin{figure}[p]
    \label{idig}
    \centering
    \renewcommand{\figurename}{Supplementary Figure}
    \includegraphics[bb = 0 0 476 220, width=\columnwidth]{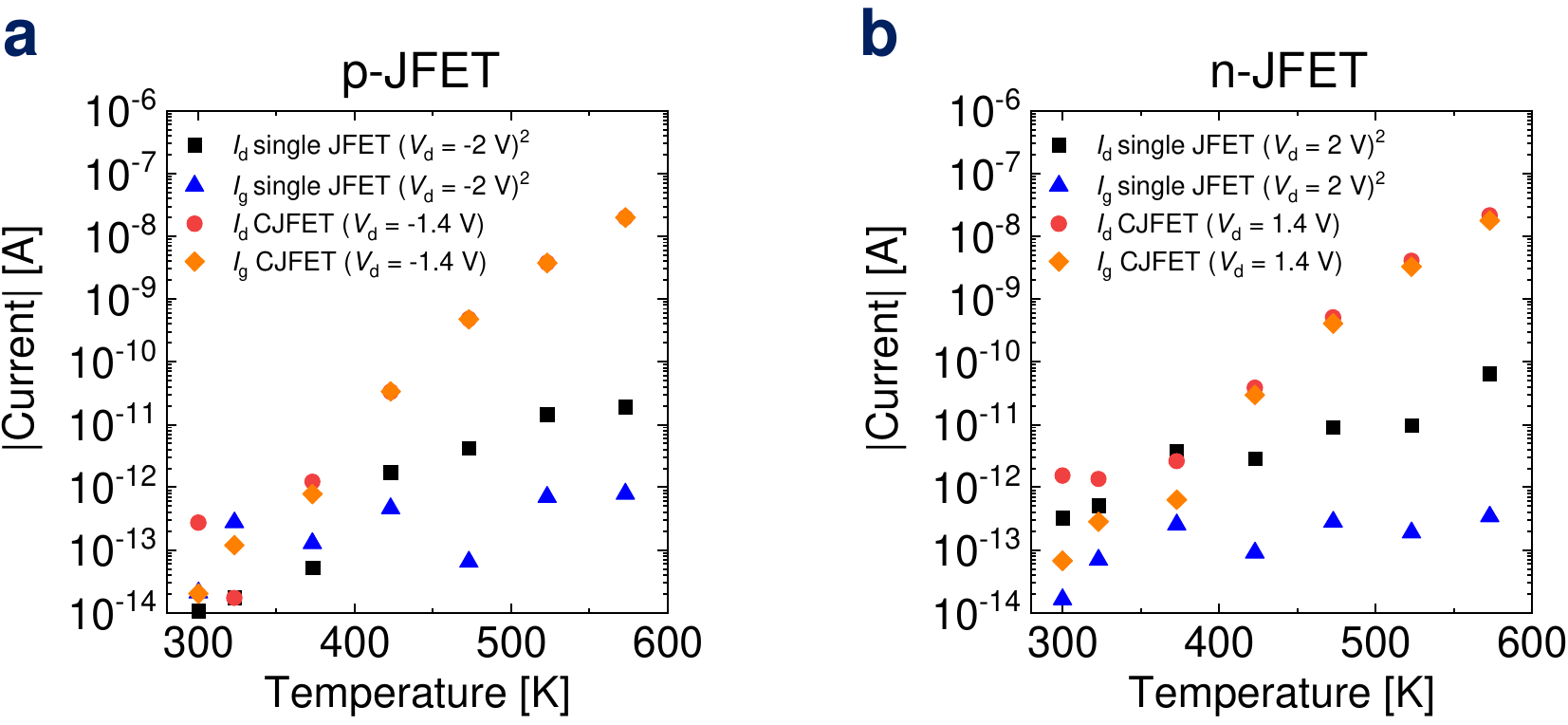}
    \caption{
        {\bf Drain and gate leakage current.} {\bf a,b}, Temperature dependence of drain and gate current without a gate supply voltage (0 V) in p-JFETs (a) and n-JFETs (b). Data obtained from the separately fabricated p- and n-JFET reported in ref. \cite{Nakajima:2019kb1} are shown.
    }
\end{figure}

\clearpage

\begin{figure}[p]
    \label{ana}
    \centering
    \renewcommand{\figurename}{Supplementary Figure}
    \includegraphics[bb = 0 0 460 458, width=\columnwidth]{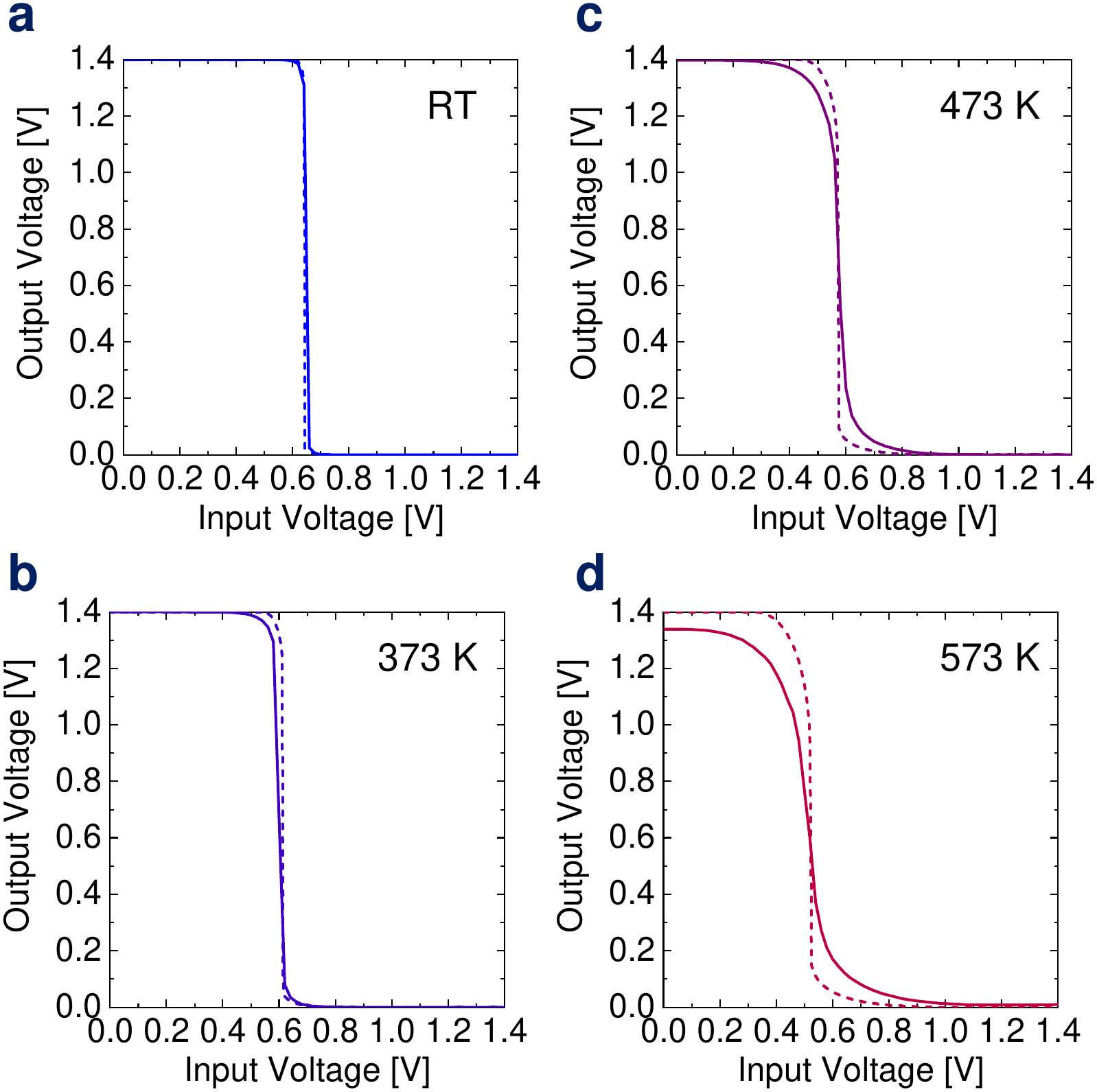}
    \caption{
        {\bf VTCs obtained from experiments and the analytical model.} {\bf a-d}, VTCs obtained from experiments (solid lines) and the analytical model (dashed lines) at RT (a), 373 K (b), 473 K (c), and 573 K (d). 
    }
\end{figure}

\setcounter{table}{0}
\renewcommand{\arraystretch}{0.5}
    \begin{table}[htbp]
      \renewcommand{\tablename}{Supplementary Table}
         \caption{Ion implantation conditions performed in this study for forming $\rm{n}^{+}$, $\rm{p}^{+}$, n, and p regions. The ion implantation was performed with the sample temperature of 300$^\circ$C for $\rm{n}^{+}$ and $\rm{p}^{+}$ regions and RT for n and p regions.}
     \begin{minipage}{.5\hsize}
       \centering
       \subcaption{$\rm{n}^{+}$ region}
    \begin{tabular}{cc}
    \hline\hline
    Energy [keV] & Dose [$\rm{cm}^{-2}$] \\
    \hline\hline
    10 & $2.50\times 10^{13}$ \\
    20 & $3.50\times 10^{13}$ \\
    35 & $8.00\times 10^{13}$ \\
    55 & $4.50\times 10^{13}$ \\
    75 & $1.50\times 10^{14}$ \\
    100 & $6.00\times 10^{13}$ \\
    130 & $2.00\times 10^{14}$ \\
    180 & $1.60\times 10^{14}$ \\
    220 & $1.75\times 10^{14}$ \\
    270 & $2.75\times 10^{14}$ \\
    360 & $3.50\times 10^{14}$ \\
    430 & $2.00\times 10^{14}$ \\
    520 & $4.50\times 10^{14}$ \\
    650 & $3.50\times 10^{14}$ \\
    700 & $5.00\times 10^{14}$ \\
    \hline
    Total dose & $3.06\times 10^{15}$  \\
    \hline\hline
    \end{tabular}
      \end{minipage}
      \hfill 
      \begin{minipage}{.5\hsize}
        \centering
      \subcaption{$\rm{p}^{+}$region}
    \begin{tabular}{cc}
    \hline\hline
    Energy [keV] & Dose [$\rm{cm}^{-2}$] \\
    \hline\hline
    10 & $2.40\times 10^{13}$ \\
    15 & $1.00\times 10^{13}$ \\
    20 & $3.30\times 10^{13}$ \\
    25 & $3.50\times 10^{13}$ \\
    40 & $1.00\times 10^{14}$ \\
    50 & $3.00\times 10^{13}$ \\
    70 & $1.00\times 10^{14}$ \\
    80 & $1.00\times 10^{14}$ \\
    120 & $3.00\times 10^{14}$ \\
    170 & $1.80\times 10^{14}$ \\
    200 & $1.80\times 10^{14}$ \\
    220 & $2.00\times 10^{14}$ \\
    250 & $7.00\times 10^{13}$ \\
    310 & $5.00\times 10^{14}$ \\
    390 & $2.50\times 10^{14}$ \\
    450 & $5.00\times 10^{13}$ \\
    470 & $5.00\times 10^{14}$ \\
    600 & $4.00\times 10^{14}$ \\
    650 & $2.50\times 10^{14}$ \\
    700 & $6.00\times 10^{14}$ \\
    \hline
    Total dose & $1.80\times 10^{15}$  \\
    \hline\hline \\
    \end{tabular}
      \end{minipage}\\
    
       \begin{minipage}{0.5\hsize} 
         \centering
     \subcaption{n region}
    \begin{tabular}{ccccc}
    \hline\hline
    Energy [keV]& Dose [$\rm{cm}^{-2}$]  \\ \hline\hline
    270 & $3.30\times 10^{11}$\\
    360 & $2.90\times 10^{11}$\\
    390 & $8.50\times 10^{10}$\\
    450 & $2.90\times 10^{11}$\\
    550 & $4.20\times 10^{11}$\\
    700 & $7.80\times 10^{11}$\\
    \hline
    Total dose & $2.20\times 10^{12}$\\
    \hline\hline
    \end{tabular}
      \end{minipage}
      \hfill
      \begin{minipage}{0.5\hsize} 
        \centering
      \subcaption{p region}
    \begin{tabular}{ccc}
    \hline\hline
    Energy [keV]& Dose [$\rm{cm}^{-2}$]  \\
    \hline\hline
    200 & $3.00\times 10^{11}$ \\
    270 & $4.10\times 10^{11}$ \\
    330 & $1.50\times 10^{11}$ \\
    360 & $2.50\times 10^{10}$ \\
    390 & $3.40\times 10^{11}$ \\
    430 & $6.00\times 10^{12}$ \\
    \hline
    Total dose & $1.83\times 10^{12}$ \\
    \hline\hline
    \end{tabular}
      \end{minipage}
    \end{table}
\clearpage
\section*{Supplementary references}

\end{document}